\documentclass[%
 aip,
rsi,%
 amsmath,amssymb,
 reprint,%
]{revtex4-1}

\usepackage{graphicx}
\usepackage{dcolumn}
\usepackage{bm}
\usepackage{nicefrac}
\usepackage{gensymb} 
\usepackage{textcomp}
\usepackage{xcolor}
\usepackage{colortbl}
\usepackage{array}
\usepackage{amsmath}
\usepackage{comment}
\newcolumntype{P}[1]{>{\centering\varraybackslash}p{#1}}

\usepackage[utf8]{inputenc}

\begin{document}

\preprint{AIP/123-QED}

\title{Transition regime in the ultrafast laser heating of solids}

\author{R. Shayduk}
\affiliation{%
European XFEL GmbH, Holzkoppel 4, D-22869 Schenefeld, Germany
}
\author{P. Gaal}
 \email{peter.gaal@ikz-berlin.de}
\affiliation{%
Leibniz-Institut f{\"u}r Kristallz{\"u}chtung, Max-Born-Str. 2, 12489 Berlin, Germany
}
\date{\today}

\begin{abstract}
Based on the phenomenological theory of heat diffusion, we show that the generated peak temperature $T_{\text{max}}$ after absorption of a laser pulse strongly depends on the pulse duration. We identify three different heat conduction regimes which can be identified via a simple parameter that depends only on the pulse duration and on material constants. The phenomenological approach is supported by numerical simulations of heat diffusion and measurements of the thermal surface expansion after transient grating excitation with 1\,ps and 10\,ns optical pulses. 
\end{abstract}

\maketitle

\section{\label{sec:level1}Introduction\protect}
Excitation and processing of materials with pulsed lasers has become a versatile tool in science and industry. For example, impulsive laser heating is used to generate shock waves for dynamic compression studies\cite{Scha2017a,ohl2015a,pepi2019a}, in photoacoustic material spectroscopy\cite{Thom1986,Ruel2015,bohj2013a} or in industrial applications, e.g., in laser ablation\cite{stee2010a,Ibra2015a}, laser cutting\cite{wetz2016a,dube2008a} or laser marking \cite{,lu2017a}. In many other experiments and applications, laser heating, although an unwanted side effect, must be considered.
Commercial pulsed laser sources today deliver pulses with a duration from few femtoseconds up to hundreds of nanoseconds. Thus, the relevant timescale for laser heating stretches at least over five orders of magnitude. In addition to the pulse duration, the light-matter interaction depends also on other parameters like the laser wavelength $\lambda$, the laser fluence and the pulse repetition rate. Often the best combination of these quantities is found in empirical studies. 

In this paper we derive a parameter to characterize the heat diffusion dynamics after absorption of a laser pulse in an opaque medium. The parameter depends only on  material constants and on the laser pulse duration and allows for a quick estimation of the generated peak temperature at the sample surface.

\section{Phenomenological theory}

In this section we briefly review the phenomenological theory of heat diffusion as described by the heat diffusion equation when applied to the case of pulsed optical excitation of semi-infinite solids. In particular we are interested in the maximum temperature $T_{\text{max}}$ at the solid surface after the optical excitation and its dependence on the duration of the exciting laser pulse. We disregard the nature of the diffusion process, i.e., whether the thermal energy is transported vie phonons\cite{lind2018a}, electrons\cite{zhou2016a} magnons\cite{uchi2008a} or other excitations. Hence, we do not apply multi-temperature models, although it is actually necessary on ultrashort timescales, i.e. for optical pulses shorter than 1\,ps. Instead we assume that all subsystems of the solid are in thermal equilibrium and compare the heat diffusion dynamics in a broad temporal range from femtoseconds to microseconds. In the following, we also neglect convection and radiative heat exchange which contribute only marginally to the heat transport in our conditions. 

The energy conservation law in our case takes the form of the heat balance equation
\begin{equation}
    c_p\rho\frac{\partial T}{\partial t}=-\nabla\cdot\mathrm{\textbf{q}}+Q,
    \label{equ:heatbalance} 
\end{equation} where $\mathrm{\textbf{q}}$  is the heat energy flux [W/m$^2$], $c_p\rho$ is the heat capacity per unit volume [J/m$^3$K], $Q$ is the external heat source [W/m$^3$] and $T$ denotes temperature [K].
In the simplest case, the heat flow at any point in the solid is determined by the direction and the magnitude of the thermal gradient as denoted in Fourier's law 
\begin{equation}
\mathrm{\textbf{q}}=-k\nabla T
    \label{equ:FourierLaw} 
\end{equation}
where $k$ [W/m$\cdot$K] is the heat conductivity coefficient. In the general case, $k$ is a tensor describing anisotropic thermal conductivity. However, as will be derived later, we restrict ourselves to one-dimensional heat transport and $k$ becomes a scalar quantity. Combination of \ref{equ:heatbalance} and \ref{equ:FourierLaw} yields the classical heat diffusion equation (HDE)
\begin{equation}
    c_p\rho\frac{\partial T(\mathbf{r},t)}{\partial t} - k\nabla^2 T(\mathbf{r},t) = Q(\mathbf{r},t)
    \label{equ:HeatEquationGeneral} 
\end{equation} that must be solved with the initial condition
\begin{equation}
    T(\mathbf{r},0)=T_0
    \label{equ:InitialConditions} 
\end{equation}

\begin{figure}
	\centering
	\includegraphics[width = 0.48\textwidth]{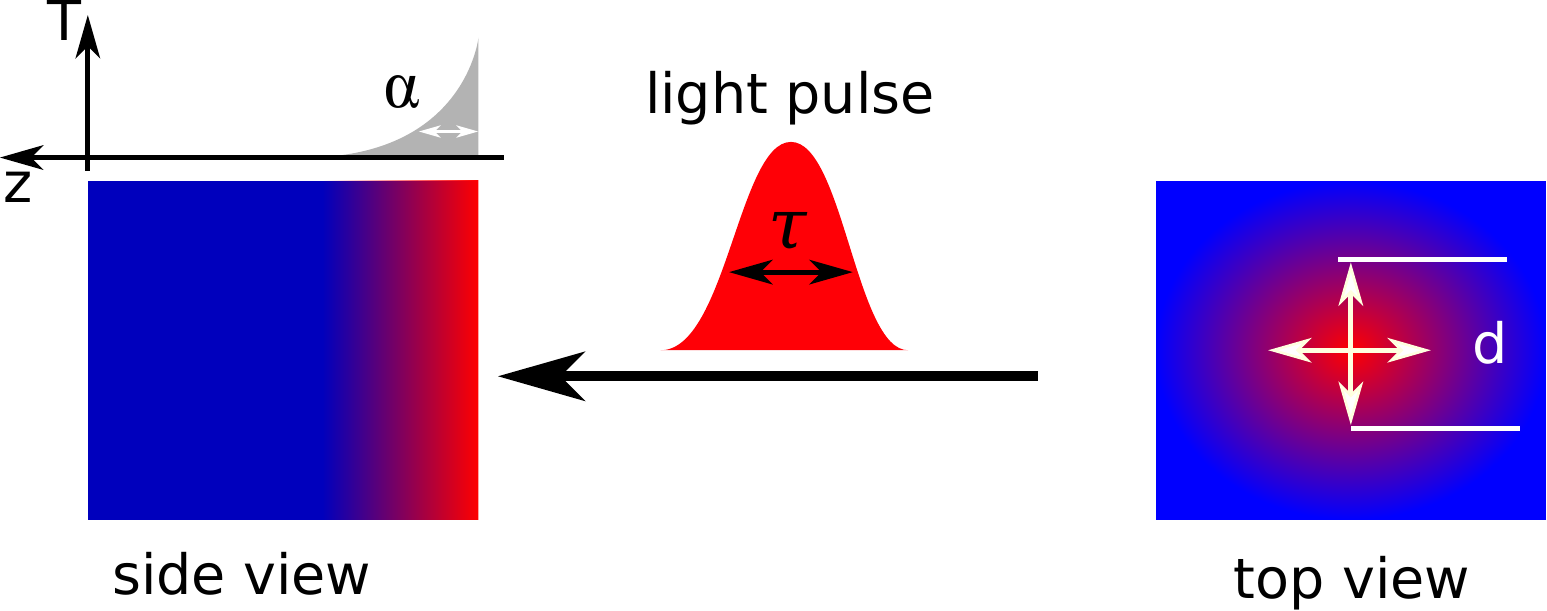}
	\caption{\textbf{Excitation Geometry}: A gaussian laser pulse with width $d$ and duration $\tau$ shape is absorbed at the sample surface. The red areas depict the initial temperature profile. The characteristic absorption length $\alpha$ is indicated in the graph.}
	\label{fig:ExcitationGeometry}
\end{figure}

The external heat source term $Q$ denotes the absorption of a laser pulse with Gaussian shape in space and time. In the approximation of short optical absorption depth $a\ll c\tau$, where $c$ is the speed of light and $\tau$ is the pulse duration, the heat source in cylindrical coordinates reads 
\begin{equation}
    Q(r,z,t)=\frac{E_0 (1-R)}{\pi^{\frac{3}{2}}\tau d^2 a}e^{-\frac{r^2}{d^2}}e^{-\frac{t^2}{\tau^2}}e^{-\frac{z}{a}}.
    \label{equ:HeatSource} 
\end{equation}
$E_0$ is the incident pulse energy [J] of the optical pulse, $\tau$ and $d$ are the $1/e$ pulse duration and beam size, respectively, $a$ is the optical absorption length, $R$ is the optical reflection coefficient. Integration over space and time of eqn.~\ref{equ:HeatSource} yields the absorbed pulse energy equal to $E_0(1-R)$. The excitation geometry is depicted in fig.~\ref{fig:ExcitationGeometry}

To retrieve the full heat diffusion dynamics one has to solve Equations~\ref{equ:HeatEquationGeneral}-\ref{equ:HeatSource}. Various more or less elaborate finite element method (FEM) solver are available for that task. Often however, a full solution of the system of equations is not necessary, for example if a specific sample geometry or a specific timescale are studied. In many applications like laser ablation or laser-based material processing it is sufficient to know the maximum of the surface temperature. 

Instead of solving the full heat diffusion problem, we derive a simple parameter which characterize the heat transport regime and which allow for a simple estimation of the maximum surface temperature $T_{\text{max}}$ after absorption of the optical pulse. This characteristic heat diffusion parameter depends only on material constants and on the duration of the exciting laser pulse and allows for a simple but accurate evaluation of the diffusion problem, even in complicated problems, e.g., thin film heterostructures.


\section{Heat diffusion in opaque media with small optical absorption depth}
\label{sec:Alpha_small}

In this section we discuss heat diffusion in a solid after localized excitation with a single laser pulse of variable duration. For optical excitation of metal targets the optical absorption coefficient $\alpha=1/a$ is large, i.e., the absorption length $a\ll d$ is much shorter than the lateral beam size $d$. Thus, on timescales 
\begin{equation}
    t\ll\tau_{\parallel}=\frac{d^2 c_p \rho}{2 k}
\end{equation} 
the heat flow is directed perpendicular to the surface because $\nabla T \approx \partial T/\partial z$ where the $z$-direction is the direction of the surface normal. The heat equation \ref{equ:HeatEquationGeneral} simplifies to
\begin{eqnarray}
    c_p\rho\frac{\partial T(r,z,t)}{\partial t} & = &k\frac{\partial^2}{{\partial z} ^2}T(r,z,t) \nonumber \\
    &+& \frac{E_0 (1-R)}{\pi^{\frac{3}{2}}\tau d^2 a}e^{-\frac{r^2}{d^2}}e^{-\frac{t^2}{\tau^2}}e^{-\frac{z}{a}},
    \label{equ:HeatEquation1D} 
\end{eqnarray}
,i.e., the lateral temperature profile is determined by the lateral profile of the exciting laser pulse energy density and not modified by heat diffusion processes on these timescales. 

In fact, $\tau_{\parallel}$ denotes the necessary diffusion time for a one dimensional, step-like temperature distribution to broaden into an error function profile of width $d$. In our problem, the parameter describes the lateral heat diffusion in the $x$-$y$-plane. Because the heat source, which is an absorbed laser pulse with a lateral Gaussian intensity profile with $\sigma_{e}=d$, already introduces lateral temperature gradients of width $d$, in-plane heat diffusion can be neglected on time scales much shorter than $\tau_{\parallel}$. For a beam size of around 1\,mm hitting Pt metal the characteristic time $\tau_{\parallel}$ is around 10\,ms.

\subsection{Excitation with ultrashort laser pulse}
\label{subsec:TauUshort}

Now we define a characteristic timescale similar to $\tau_{\parallel}$ but for heat diffusion in perpendicular direction
\begin{equation}
    \tau_{\perp}=\frac{a^2 c_p \rho}{2 k}
    \label{equ:defTau2}
\end{equation} 
In analogy to $\tau_{\parallel}$, this parameter describes the diffusion of a step-like heat distribution into a smooth Gaussian heat profile. For ultrashort optical pulses with duration
\begin{equation}
\tau\ll\tau_{\perp} 
\label{equ:shortpulselimit}
\end{equation}
the heat diffusion term in eqn.~\ref{equ:heatbalance} is negligible compared to the source term $Q$ on the timescale of the pulse duration.
The rate of temperature change only depends on the external heat source i.e., on the absorbed laser pulse. Hence, the temperature field at a time $t$ is determined by the integral
\begin{equation}
    T(r,z,t) = \frac{E_0 (1-R)}{c_p\rho\pi^{\frac{3}{2}}\tau d^2 a}e^{-\frac{r^2}{d^2}}e^{-\frac{z}{a}}\int_{-\infty}^{t}e^{-\frac{t'^2}{\tau^2}}\mathrm{d}t' 
    \label{equ:SolutionNoHeatDiffusion} 
\end{equation}
The maximum temperature rise $T_{\text{max}}$ occurs at the surface in the center of the laser pulse
\begin{equation}
    T(0,0,\infty) = \frac{E_0 (1-R)}{c_p\rho\pi d^2 a}= \frac{F}{c_p\rho a} = T_{\mathrm{max}}
    \label{equ:PeakTemperature} 
\end{equation}
with the absorbed fluence [J/m$^2$]
\begin{equation}
    F=\frac{E_0(1-R)}{\pi d^2}
    \label{equ:Fluence} 
\end{equation}

Thus, in the limit $\tau\ll\tau_{\perp}$ where the heat diffusion term can be neglected, the maximum temperature change $T_{\text{max}}$ is independent of the duration of the optical pump pulse.

\subsection{Excitation with long laser pulses}
\label{subsec:TauLong}
Again we use eqn.~\ref{equ:defTau2} to define the regime of long pulse duration:
\begin{equation}
\tau \gg \tau_{\perp}
\label{equ:longpulse}
\end{equation}
which may also be expressed as
\begin{equation}
a \ll \frac{2k\tau}{c_p\rho}
\end{equation} 
In the regime of long pulse duration the optical absorption length is much shorter than the characteristic heat diffusion length. On timescales $t\gtrapprox\tau$ the heat diffusion term in eqn.~\ref{equ:HeatEquation1D} dominates. This limit is thoroughly discussed in literature\cite{bech1975a,anto2013a}. Here, we rely on the solution presented by Bechtel\cite{bech1975a}. The heat diffusion problem is solved by replacing the inhomogeneous heat equation~\ref{equ:HeatEquation1D} by the homogeneous heat equation with the heat source defined as the boundary condition:
\begin{eqnarray}
-k\frac{\partial T}{\partial z}(r,0,t)&=&\frac{F}{\sqrt{\pi}\tau a}e^{-\frac{r^2}{d^2}}e^{-\frac{t^2}{\tau^2}}\int e^{-\frac{z}{a}}\mathrm{d}z \nonumber \\
    & = &\frac{F}{\sqrt{\pi}\tau}e^{-\frac{r^2}{d^2}}e^{-\frac{t^2}{\tau^2}}
\label{equ:SolBechtel}
\end{eqnarray}
A detailed discussion of Bechtel's solution is presented by Shayduk et al.\cite{shay2016a} and compared to experimental data in the context of pulsed nanosecond laser heating of a Platinum surface. The maximum of the temperature T$_{\mathrm{max}}$\,[K] is reached at $t\approx 0.54\cdot\tau$ and reads
\begin{equation}
T_{\mathrm{max}} \approx 2.15 \frac{F}{\pi \sqrt{k \rho c_p \tau}}
\label{equ:TmaxShayduk}
\end{equation}

In order to compare the limit of excitation with a short and a long optical pump pulse, we normalize the peak temperature [c.f. eqn.~\ref{equ:TmaxShayduk}] to the result of the short pulse regime [c.f. eqn.~\ref{equ:PeakTemperature}]. Assuming similar pulse fluencies in both cases, i.e., a similar absorbed energy $E_{0}(1-R)$, we find that the temperature rise at the sample surface reads
\begin{equation}
T^{\text{surf}}_{\text{max}} \approx \sqrt{ \frac{\tau_{\perp}}{\tau}}
\label{equ:TmaxNormalized}
\end{equation} 
This solution is experimentally verified for long excitation pulses\cite{shay2016a}. This expression becomes invalid for $\tau\leq\tau_{\perp}$ because  $T^{\text{surf}}_{\text{max}}>1$. Thus, for an intermediate pulse duration
\begin{equation}
\tau \approx \tau_{\perp}
\label{equ:TransitionTime}
\end{equation}
there is a transition regime, where neither the heat diffusion term [c.f. sec.~\ref{subsec:TauUshort}] can be neglected, nor the surface absorption boundary condition [c.f. eqn.~\ref{equ:SolBechtel}] applies. Instead the full heat diffusion equation \ref{equ:HeatEquation1D} must be solved.

\subsection{Intermediate transition regime}
\label{sec:IntermediateRegime}
\begin{figure}
  \centering
  \includegraphics[width = 0.48\textwidth]{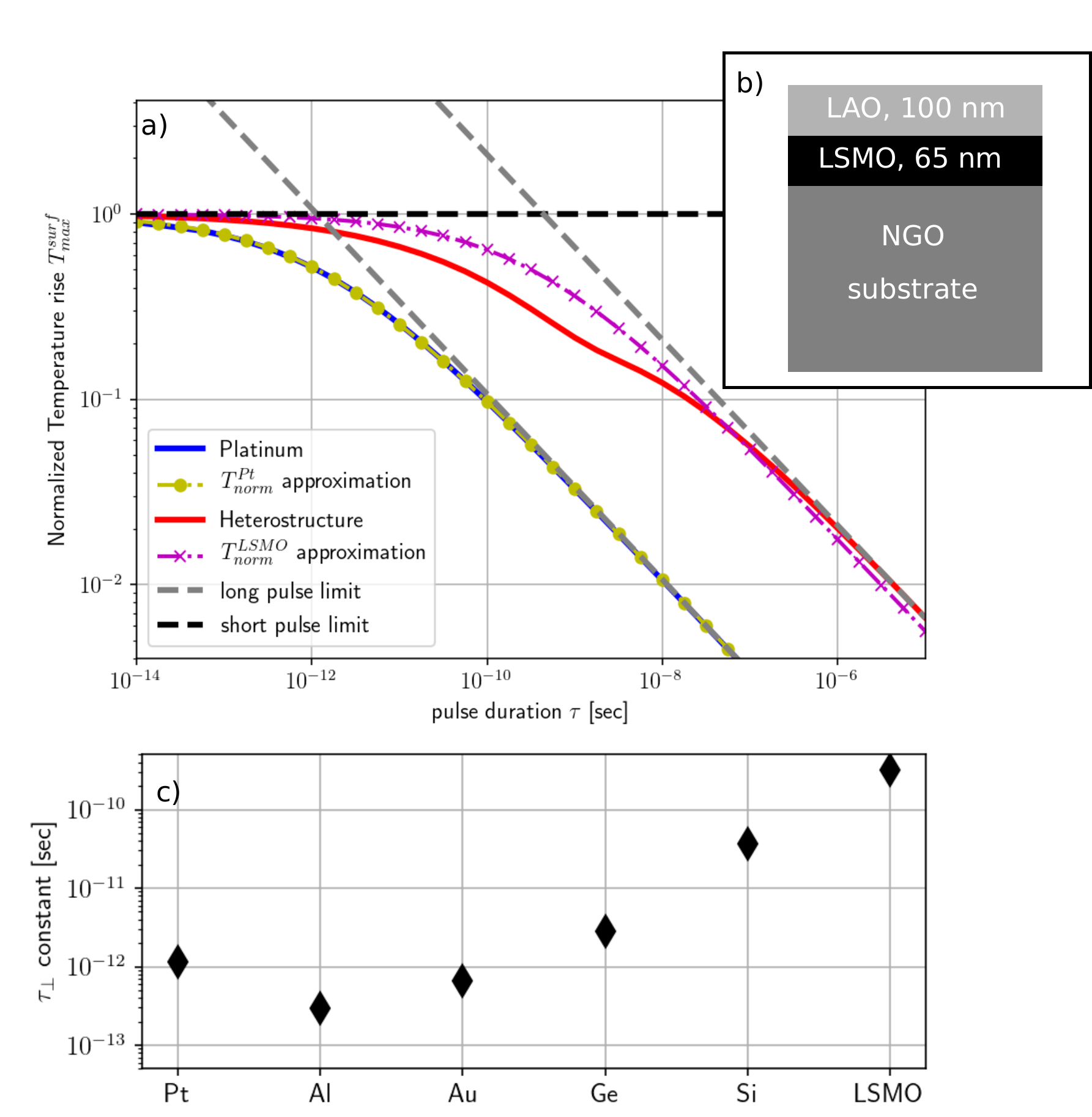}
  \caption{\textbf{FEM Simulation results}: a) Simulated peak temperature vs. duration of the excitation pulse. The solid lines denote results for Platinum (Pt, blue) and for the LaAlO$_{3}$ (LAO) on (La,Sr)MnO$_{3}$ (LSMO) heterostructure grown on NdGaO$_{3}$ (NGO) substrate (red). The sample is depicted in b). c) Calculated values for the transition time constant \textbf{$\tau_{\perp}$} Note that for Sillicon (Si) and Germanium (Ge) we used the optical absorption length $a$ at a wavelength of 400\,nm, while for Pt, Aluminum (Al), Gold (Au) and for the heterostructure sample we used $a$ at a wavelength of 800\,nm.}
  \label{fig:Simulation}
\end{figure}

In order to characterize the transition regime, we solve the full heat diffusion equation [c.f. eqn.~\ref{equ:HeatEquationGeneral}] numerically with the finite element solver COMSOL. We implement a source term with variable pulse duration in the range of 10$^{-14}$\,sec to 10$^{-5}$\,sec. The absorbed pulse energy in the simulation was 2.4\,mJ/cm$^{2}$. It is kept constant for all pulse durations and the simulation was performed assuming a Gaussian footprint on the sample surface. The simulation yields a spatiotemporal temperature field which we evaluate by extracting the maximum temperature rise and normalize it to eqn.~\ref{equ:PeakTemperature}, i.e., $T_{\text{max}}$ in the short pulse limit. The simulations were performed for bulk Platinum (Pt) and for a thin film heterostructure sample which consists of a top layer of optically transparent LaAlO$_{3}$ (LAO), an optically opaque layer of La$_{0.7}$Sr$_{0.3}$MnO$_{3}$ (LSMO) and a transparent NdGaO$_{3}$ (NGO) substrate. The layer thickness was 100\,nm and 65\,nm, respectively, the structure is sketched in fig.~\ref{fig:Simulation}~b). Simulation parameters are given in tab.~\ref{tab:simulation} and experimental data measured on a similar heterostructure is shown in the experimental section~\ref{sec:Experiments}.

The normalized peak temperature $T^{\text{surf}}_{\text{max}}$ for different duration of the optical pump pulse in both samples is depicted by the solid lines in fig.~\ref{fig:Simulation}~a). The gray dashed line depicts the normalized solution in the long time limit [c.f. eqn.~\ref{equ:TmaxNormalized}]. The black dashed line marks the maximum normalized temperature $T^{\text{surf}}_{\text{max}}=1$. First, we discuss the blue solid line, which depicts the temperature dependence for bulk Pt. According to eqn.~\ref{equ:TransitionTime} and eqn.~\ref{equ:defTau2} we expect the transition regime for a pulse duration of $\tau_{\perp}\approx1.15$\,ps. For shorter excitation pulses, $T^{\text{surf}}_{\text{max}}$ approaches unity and excitation pulses longer than 100\,ps $T^{\text{surf}}_{\text{max}}$ follows the gray dashed line which depicts the long pulse limit.

\begin{table}
\centering
\begin{tabular}{|c|c|c|c|c|} 
  \hline
  & $c_{p}$ & $k$  & $\rho$ & $a$ \\
  & [J kg$^{-1}$K$^{-1}$] & [W m$^{-1}$ K$^{-1}$] & [g m$^{-3}$] & [nm$^{-1}$] \\
 \hline
 \hline
 LAO & 438 & 12 & 6.52 & $\infty$ \\
 LSMO & 598 & 12 & 6.6 & 44 \\
 NGO & 344 & 7 & 7.56 & $\infty$ \\
 Pt & 130 & 71.6 & 21.45 & 8 \\
 \hline
 \end{tabular}
 \caption{\textbf{Simulation parameters for the numerical solution of the HDE:} $c_{p}$, $k$, $\rho$, $a$ denote the specific heat, thermal conductivity, mass density and optical absorption coefficient, respectively}
\label{tab:simulation}
\end{table}

Temperature dependence on the pulse duration for the heterostructure sample is shown by the red solid line. We estimate the transition time by using only material parameters of the absorbing layer, i.e., (La,Sr)MnO$_{3}$. With the values given in tab.~\ref{tab:simulation} we find $\tau_{\perp}\approx320$\,ps, i.e., more than 100 times longer than that for Pt.

Fig.~\ref{fig:Simulation} shows that the transition regime stretches over several orders of magnitude. The short pulse limit yields accurate estimations of the peak temperature if femtosecond optical pulses are used for the excitation. Deviation from the short time limit may occur even for picosecond excitation pulses. The long time limit is valid only for excitation pulses with nanosecond or even microsecond duration. 

The $\tau_{\perp}$ constant is a material parameter, which describes heat transport after optical excitation of a solid with a single laser pulse. In general, the peak temperature after absorption of a laser pulse shorter than 1\,ps can be calculated in the short time limit using eqn.~\ref{equ:PeakTemperature} in any material. As stated below, one might have to consider multi-temperature models on these ultrashort timescales. Metals tend to have small values of $\tau_{\perp}$, due to the high thermal conductivity and the short optical penetration length. In other materials, e.g., semiconductors, $\tau_{\perp}$ may become much longer. As shown by the example of the heterostructure sample, the transition regime may extend up to a pulse duration of 100\,ns. Commercially available pulsed lasers rarely provide longer pulses. Thus, except for femtosecond laser excitation, the peak temperature after laser excitation must be evaluated in the transition regime. Examples for transition time constant $\tau_{\perp}$ of different materials are given in fig.~\ref{fig:Simulation}~c).  

In addition to estimating the transition pulse duration, $\tau_{\perp}$ may also be used to approximate the temperature rise after optical excitation across the whole range of pulse durations. The normalized approximated temperature rise is given by
\begin{equation}
T_{approx}^{norm}=\frac{1}{\sqrt{1+\tau/\tau_{\perp}}}
\label{equ:Tapprox}
\end{equation}
and results of the approximation are shown by the dash-dotted lines marked for clarity with yellow bullets and magenta crosses in fig.~\ref{fig:Simulation}~a), respectively. The approximated temperature rise matches the simulated result for Pt. For the heterostructured sample we observe a deviation since the temperature distributes across several materials. A better representation could be obtained with an effective medium value for $\tau_{\perp}$ of the heterostructure. We believe that fig.~\ref{fig:Simulation} together with eqn.~\ref{equ:Tapprox} provide an easy-to-use recipe to retrieve the peak temperature after laser excitation with arbitrary pulse duration. A reliable approximation can be derived directly from material parameters of the sample by calculating $\tau_{\perp}$. In particular, it is often not necessary to solve the complex heat diffusion problem. 

We want to point out again that our calculations are based on a phenomenological theory. For excitation pulses shorter than 2\,ps we generally recommend using multi-temperature models to retrieve peak temperatures $T_{\text{max}}$. In the short pulse limit temperatures may deviate from predictions of our model\cite{lin2008a}. In materials with slow electron-phonon coupling, such as Au\cite{whit2014a}, the optical absorption depth may no longer determine the initial temperature profile after pulse absorption due to ballistic electron transport\cite{zhou2017a}. However, as soon as thermal equilibrium of electrons and phonons is reached, our phenomenological theory applies.

However, our phenomenological model yields a good first estimate of the heat transport regime based solely on material parameters and on the laser pulse duration. In particular, we showed that the intermediate transition regime stretches over a broad range of pulse duration. In fact, heat transport in many industry-relevant materials happens in the transition regime\cite{stee2010a,lawr2015a,sale2014a,hans2010a}. The commonly applied heat diffusion equation \ref{equ:HeatEquationGeneral} with a surface heat source derived by Bechtel\cite{bech1975a} overestimates the surface temperature in the intermediate regime, as shown by the dashed and solid lines in fig.~\ref{fig:Simulation}.

\section{Experimental Data}
\label{sec:Experiments}
We employ time-resolved x-ray diffraction to studying thermal diffusion in solids by detecting the transient lattice temperature via the lattice expansion after absorption of a optical excitation pulse.\cite{shay2011,navi2014a,navi2011a} In a similar way, thermal expansion of a laser-excited solid surface can be detected with ultrahigh precision using the excitation of so-called transient gratings (TG) which are probed by time-resolved x-ray reflectivity (TRXRR).\cite{sand2017a} 

The excitation and TRXRR probing of TGs is described in detail elsewhere\cite{pude2019a, sand2017b} and only the main features are summarized here. TGs are excited by overlapping two laser pulses on the surface of a sample. Interference of the two optical beams results in a spatial modulation with a period $\Lambda$ of few micrometers. The spatial modulation of the excitation is inscribed as thermally expanded surface in the sample. An x-ray pulse impinging the sample under grazing incidence angles is diffracted from the periodically expanded surface due to a momentum transfer $\vec{k}_{\text{out}} - \vec{k}_{\text{in}}=\vec{q}_{\perp}\pm\frac{2\pi}{\Lambda}$. $\vec{k}_{\text{in}}$,$\vec{k}_{\text{out}}$ and $\vec{q}_{\perp}$ denote the wavevectors of the incident and diffracted beam and the recoil momentum of the specular reflection at the surface, respectively. It has been shown that the x-ray intensity of the first diffraction order from the TG is a precise measure for the amplitude of the surface deformation.\cite{sand2017b} Thus, by measuring the first order diffracted intensity, we can determine the thermal expansion of the laser-excited surface. As explained below, our setup allows to select only two distinctly different pulse durations, namely 1\,ps and 10\,ns. Thus, we are not able to reconstruct the nonlinear relation of pulse duration and generated peak temperature. In consequence, we do not convert the volumetric thermal expansion to a transient temperature in the excited film. Instead, we use the volumetric thermal expansion as an indicator for the absorbed energy in the sample. Although this is only an indirect signature of the generated temperature profile, we believe that the experimental data is an illustrative example for the theory derived in the previous sections. 

The experiments were performed at the ID09 beamline at the European Synchrotron ESRF.\cite{wulf2003,camm2009} The beamline is equipped with a commercial Ti:Sapphire laser system (Legend, Coherent Inc.) that delivers optical pulses with a wavelength $\lambda$=800\,nm, a pulse energy of $E$=2\,mJ and a duration of $\tau_{p}$=1\,ps at a repetition rate of $f$=1\,kHz. The laser system does not provide an adjustment of the pulse duration. 
However, the cavity of the regenerative amplifier emits pulsed amplified stimulated emission (ASE) by itself, which has a duration of $\tau_{p}\approx$10\,ns.\cite{ivan2003a} Thus, by blocking the cavity seed pulse and by bypassing the compressor stage of the amplifier \cite{ster1998a,lin2004a} the laser emits 10\,ns optical pulses at similar energies, wavelength and repetition rates as in the short-pulse, seeded mode.

\begin{figure}
  \centering
  \includegraphics[width = 0.48\textwidth]{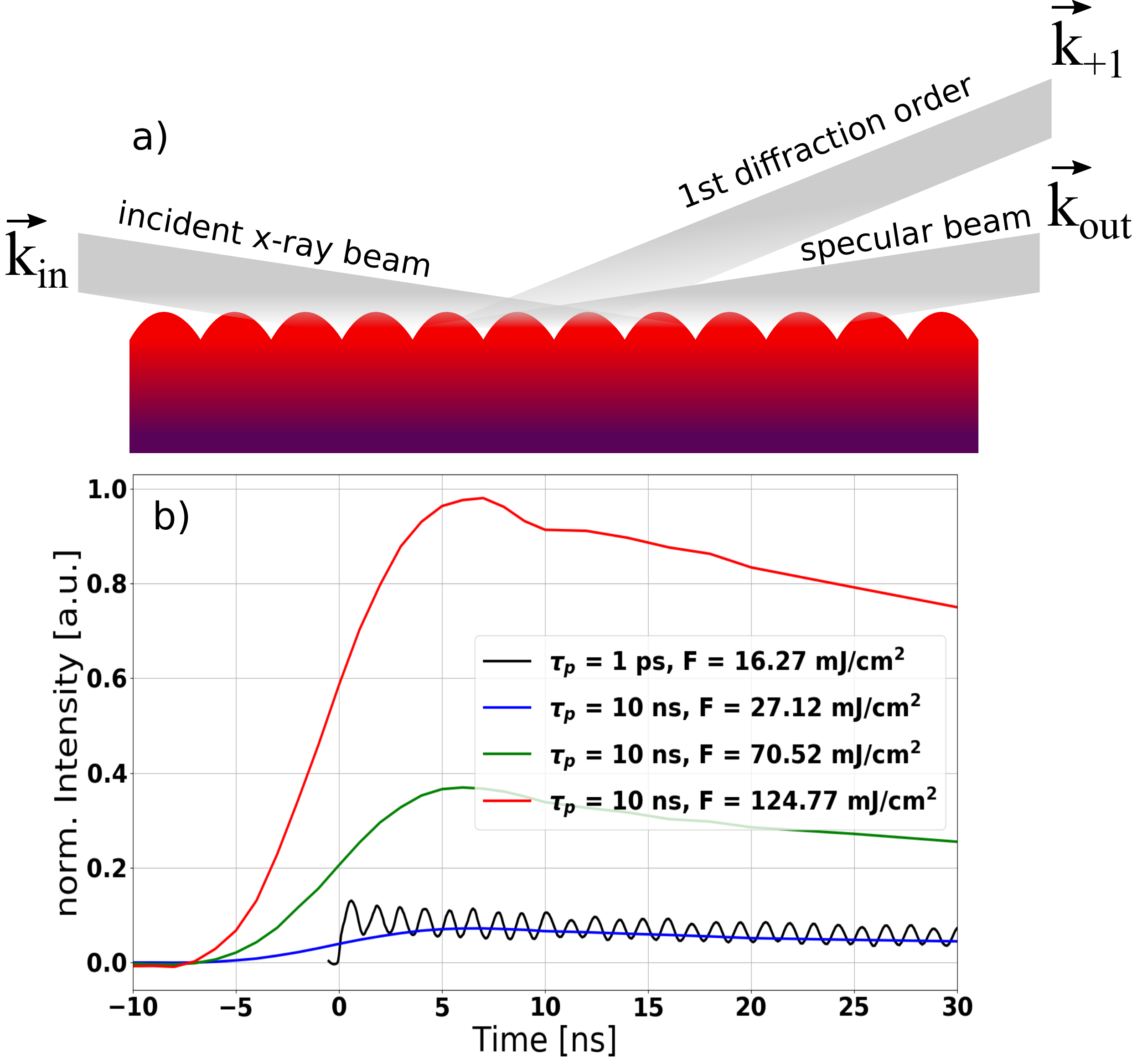}
  \caption{\textbf{Experimental Data:} a) X-ray diffraction from a periodically distorted sample after transient grating (TG) laser excitation. b) Experimental data measured after excitation with 1\,ps (black) and 10\,ns (blue, green and red) optical pulses.}
  \label{fig:data}
\end{figure}

TRXRR measurements of laser excited TGs were performed using $\tau_{p}$=1\,ps and $\tau_{p}$=10\,ns laser pulses. The sample was a LAO/LSMO/NGO heterostructure as described in sec.~\ref{sec:IntermediateRegime} and sketched in fig.~\ref{fig:Simulation}. The damage threshold of the sample after excitation with femtosecond optical pulses is $\approx$30\,mJ/cm$^{2}$. Using eqn.~\ref{equ:PeakTemperature} we find $T_{max}=1730$\,K. 
Experimental data for different excitation fluencies is shown in Fig.~\ref{fig:data}. The black curve is measured with fs excitation pulse and a fluence of 16.27\,mJ/cm$^{2}$ [$T_{max}=950$\,K]. The data exhibits characteristic oscillations from the propagation of impulsively excited coherent surface acoustic waves.\cite{sand2017a,pude2019a} 

Measurements with the 10\,ns excitation pulse and with fluencies of 27\,mJ/cm$^{2}$, 70\,mJ/cm$^{2}$ and 125\,mJ/cm$^{2}$ are depicted by the blue, green and red line in Fig~\ref{fig:data}. The latter fluence is clearly above the damage threshold fluence of the short pulse excitation. Using eqn.~\ref{equ:TmaxNormalized} we find peak temperatures of 270\,K, 700\,K and 1250\,K. The simulated peak temperature shown in fig.~\ref{fig:Simulation} is even lower because the 10\,ns excitation pulse still falls in the transition regime of laser heating. However, noticeable heat transport from the excited volume to adjacent layers occurs during the absorption of the 10\,ns optical pulse. This protects the excited volume from overheating and allows higher absorbed fluencies. The accumulated thermal expansion in the sample leads to the higher deformation and thereby, higher diffracted intensity observed in the experiment. We will discuss the experimental data in more detail in a separate publication that is currently in preparation.

\section{Application examples of the $\tau$-parameter}
\label{sec:Examples}
In this section we want to discuss the $\tau$-parameter, that was derived in sec.~\ref{sec:Alpha_small}, in the context of practical examples. As stated before, the expression 
\begin{equation}
\tau=\frac{d^{2}c_{p}\rho}{2k}
\label{equ:tauParameter}
\end{equation}
defines the timescale where the temperature distribution in a finite volume is equally determined by the source and by the diffusion term in eqn.~\ref{equ:HeatEquationGeneral}. Although we have derived the theory to describe the absorption of a single laser pulse, real-world experiments generally employ laser that emit pulses at a specific repetition rate. Therefore, in the first example, we discuss accumulated heat by femtosecond optical excitation of a solid heterostructure at high repetition rates, as described by Reinhardt et al.\cite{rein2016a}. The sample was a 94\,nm thin SrRuO$_{3}$ (SRO) metallic film on a transparent SrTiO$_{3}$ (STO) substrate. It was excited by optical pulses with a duration of 250\,fs at a repetition rate of 208\,kHz and an average power of 1.2\,W.  Receprocal space maps at the (002) STO Bragg reflex revealed a strong sample deformation due to accumulated heat under these excitation conditions. 

In order to apply the $\tau$-parameter to this experiment, we solve eqn.~\ref{equ:tauParameter} for the characteristic diffusion length $d=2\sqrt{\frac{\tau k}{c_{p}\rho}}\approx 8\,\mu m$ where $\tau=1/f_{rep}$=5\,$\mu$s now denotes the interval between excitation pulses and $c_{p}=733$ [J kg$^{-1}$K$^{-1}$], $k=12$ [W m$^{-1}$ K$^{-1}$] and $\rho=5120$ [g m$^{-3}$] denote material parameters for STO. The accumulated temperature per excitation pulses reads\cite{rein2016a}
\begin{equation}
\Delta T=\frac{F}{c_{p}\rho \cdot d} = 1.5\,[K]
\end{equation}
where $F$ denotes the fluence of the excitation pulse. Thus, within the characteristic diffusion length $d$ the temperature between two laser excitation rises by 1.5\,K.  Note that the experiment was performed at a repetition rate of 208\,kHz. Thus, the accumulation leads to significant static heating of the sample in the excitation region. This example demonstrates that the $\tau$-parameter can be used to easily estimate heat accumulation effects in all-optical pump-probe experiments.

The second example we discuss is a Si monochromater for hard x-rays, i.e. a standard optical element at Synchrotron and Free Electron Laser (FEL) beamlines. Typically broadband x-ray pulses emitted from the insertion device impinge the monochromator at the Bragg angle. As an example we discuss the (111) reflex of Si at 10\,keV. Except for a very narrow spectral portion ($\frac{\Delta E}{E_{0}}=10^{-4}$), which is diffracted towards the experiment, the impinging photons are absorbed in the monochromator within $d=27.85$\,$\mu$m\cite{step2004a}. 
By directly applying eqn.~\ref{equ:tauParameter} with material parameters for Si, we find a characteristic timescale of $\tau=55\,\mu$s. Thus, only after this time a significant amount of thermal energy is transported away form the absorption volume. This estimation makes it immediately clear that radiation induced thermal deformation is a major concern in the design of x-ray optical elements\cite{zhan2013a}. These issues will become even more important in the future with the advent of high brilliance FELs working at high repetition rates\cite{cart2016a} or with the newly installed low-emittance 4th generation synchrotron sources\cite{schr2018a}.

\section{Conclusion}
\label{sec:conclusion}
In conclusion we have identified three regimes of heat diffusion after laser excitation of opaque solids where the generated peak temperature $T_{\text{max}}$ depends differently on the pulse duration: for ultrashort excitation pulses, the heat diffusion term in eqn.~\ref{equ:HeatEquationGeneral} can be neglected and $T_{\text{max}}$ becomes independent of the pulse duration. For long excitation pulses, the heat diffusion term dominates and $T_{\text{max}}$ becomes proportional to $\tau^{-\nicefrac{1}{2}}$. This solution has been previously derived\cite{bech1975a,shay2016a} and is commonly used to determine $T_{\text{max}}$ after laser excitation. However, in the intermediate regime, $T_{\text{max}}$ must be calculated by fully solving eqn.~\ref{equ:HeatEquationGeneral}. We introduce the parameter $\tau_{\perp}$ [c.f.~eqn.~\ref{equ:defTau2}] which depends solely on material constants and on the duration of the excitation pulse and which allows to asses the relevant heat transport conditions in a given experiment. By comparison with FEM simulations we find that the long pulse limit typically overestimates the surface temperature. With the exception of metals, the intermediate regime applies in most practical cases, where excitation pulses with a duration between 10\,ps and 50-100\,ns are used. The derived theory is illustrated by measurements of the volumetric thermal expansion of a sample after excitation with optical pulses of variable pulse duration.

\textbf{Acknowledgment}

We acknowledge help during the experiment by J.E. Pudell and M. Herzog from Universit\"at Potsdam and by M. Sander, M. Levantino and M. Wulff from the European Synchrotron ESRF. We thank D. Pf\"utzenreuter and J. Schwarzkopf from Institut f\"ur Kristallz\"uchtung for providing the sample. We further acknowledge valuable discussions with Vedran Vonk and Uta Hejral from Desy NanoLab. Finally, we acknowledge financial support from BMBF via grant 05K16GU3


%

\end{document}